\documentclass{aastex62}

\received{}
\revised{}
\accepted{}
\submitjournal{ApJ}

\shorttitle{Interferometric Observations of Cyanopolyynes toward the G28.28$-$0.36}
\shortauthors{Taniguchi et al.}

\begin{document}

\title{Interferometric Observations of Cyanopolyynes toward the G28.28$-$0.36 High-Mass Star-Forming Region}

\correspondingauthor{Kotomi Taniguchi}
\email{kotomi.taniguchi@nao.ac.jp}

\author{Kotomi Taniguchi}
\altaffiliation{Research Fellow of Japan Society for the Promotion of Science}
\affiliation{National Astronomical Observatory of Japan, Osawa, Mitaka, Tokyo 181-8588, Japan}

\author{Yusuke Miyamoto}
\affiliation{National Astronomical Observatory of Japan, Osawa, Mitaka, Tokyo 181-8588, Japan}

\author{Masao Saito}
\affiliation{National Astronomical Observatory of Japan, Osawa, Mitaka, Tokyo 181-8588, Japan}
\affiliation{Department of Astronomical Science, School of Physical Science, SOKENDAI (The Graduate University for Advanced Studies), Osawa, Mitaka, Tokyo 181-8588, Japan}

\author{Patricio Sanhueza}
\affiliation{National Astronomical Observatory of Japan, Osawa, Mitaka, Tokyo 181-8588, Japan}

\author{Tomomi Shimoikura}
\affiliation{Department of Astronomy and Earth Sciences, Tokyo Gakugei University, Nukuikitamachi, Koganei, Tokyo 184-8501, Japan}

\author{Kazuhito Dobashi}
\affiliation{Department of Astronomy and Earth Sciences, Tokyo Gakugei University, Nukuikitamachi, Koganei, Tokyo 184-8501, Japan}

\author{Fumitaka Nakamura}
\affiliation{National Astronomical Observatory of Japan, Osawa, Mitaka, Tokyo 181-8588, Japan}
\affiliation{Department of Astronomical Science, School of Physical Science, SOKENDAI (The Graduate University for Advanced Studies), Osawa, Mitaka, Tokyo 181-8588, Japan}

\author{Hiroyuki Ozeki}
\affiliation{Department of Environmental Science, Faculty of Science, Toho University, Miyama, Funabashi, Chiba 274-8510, Japan}



\begin{abstract}

We have carried out interferometric observations of cyanopolyynes, HC$_{3}$N, HC$_{5}$N, and HC$_{7}$N, in the 36 GHz band toward the G28.28$-$0.36 high-mass star-forming region using the Karl G. Jansky Very Large Array (VLA) Ka-band receiver.
The spatial distributions of HC$_{3}$N and HC$_{5}$N are obtained.
HC$_{5}$N emission is coincident with a 450 $\mu$m dust continuum emission and this clump with a diameter of $\sim 0.2$ pc is located at the east position from the 6.7 GHz methanol maser by $\sim 0.15$ pc. 
HC$_{7}$N is tentatively detected toward the clump.
The HC$_{3}$N : HC$_{5}$N : HC$_{7}$N column density ratios are estimated at 1.0 : $\sim 0.3$ : $\sim 0.2$ at an HC$_{7}$N peak position.
We discuss possible natures of the 450 $\mu$m continuum clump associated with the cyanopolyynes.
The 450 $\mu$m continuum clump seems to contain deeply embedded low- or intermediate-mass protostellar cores, and the most possible formation mechanism of the cyanopolyynes is the warm carbon chain chemistry (WCCC) mechanism.
In addition, HC$_{3}$N and compact HC$_{5}$N emission is detected at the edge of the 4.5 $\mu$m emission, which possibly implies that such emission is the shock origin.
\end{abstract}

\keywords{astrochemistry --- ISM: individual objects (G28.28$-$0.36) --- ISM: molecules --- stars: formation}



\section{Introduction} \label{sec:intro}

Cyanopolyynes (HC$_{2n+1}$N, $n=1,2,3,...$) are one of the representative carbon-chain species.
In low-mass star-forming regions, carbon-chain molecules are known as early-type species; they are abundant in young starless cores and deficient in star-forming cores \citep[e.g.,][]{1992ApJ...392..551S, 2009ApJ...699..585H}.
In contrast to the general picture, cyanoacetylene (HC$_{3}$N), the shortest member of cyanopolyynes, is detected from various regions such as infrared dark clouds \citep[IRDCs; e.g.,][]{2012ApJ...756...60S}, molecular outflows \citep{1997ApJ...487L..93B}, protoplanetary disks \citep{2015Natur.520..198O, 2018ApJ...857...69B}, and comets \citep[e.g.,][]{2011ARA&A..49..471M} and it is interesting to trace cyanopolyyne chemistry for better understanding of the molecular evolution during star/planet formation process.
Cyanopolyynes attract astrobiological as well as astrochemical interests.
Since they contain the nitrile bond (--C$\equiv$N), cyanopolyynes have been suggested as possible intermediates in the synthesis of simple amino acids \citep[e.g.,][]{2017A&A...605A..57F, 2018arXiv180409210C}.

Saturated complex organic molecules (COMs), consisting of more than six atoms with rich hydrogen atoms, are abundant around protostars.
Such chemistry is known as hot core in high-mass star-forming regions and hot corino in low-mass star-forming regions. 
In addition to hot corino, around a few low-mass protostars, carbon-chain molecules are formed from CH$_{4}$ evaporated from dust grains, which is known as warm carbon chain chemistry \citep[WCCC; e.g.,][]{2013ChRv..113.8981S}.

Progress in observational studies of carbon-chain molecules in high-mass star-forming regions has been slower, compared to low-mass star-forming regions.
Regarding hot cores, HC$_{5}$N has been detected in chemically rich sources, Orion KL \citep{2013A&A...559A..51E} and Sgr B2 \citep{2013A&A...559A..47B}, while only a tentative detection of HC$_{7}$N in Orion KL was reported \citep{2015A&A...581A..71F}.
\citet{2009MNRAS.394..221C} performed a chemical network simulation and suggested that cyanopolyynes could be formed in a hot core from C$_{2}$H$_{2}$ evaporated from grain mantles.
Motivated by the chemical network simulation, \citet{2014MNRAS.443.2252G} carried out survey observations of HC$_{5}$N toward 79 hot cores associated with the 6.7 GHz methanol masers and reported its detection in 35 sources.
However, the association with the maser is questionable, because they used a large beam (0.95\arcmin) and a low-excitation energy line ($J=12-11$; $E_{\rm u}/k = 10.0$ K), which can be excited even in dark clouds.

\citet{2017ApJ...844...68T} carried out observations of long cyanopolyynes (HC$_{5}$N and HC$_{7}$N) toward four massive young stellar objects, where \citet{2014MNRAS.443.2252G} had reported the HC$_{5}$N detection, using the Green Bank 100-m and the Nobeyama 45-m radio telescopes, and detected high-excitation energy lines ($E_{\rm u}/k \approx 100$ K) of HC$_{5}$N.
The detection of such lines means that HC$_{5}$N exists at least in the warm gas, not in cold molecular clouds ($T_{\rm {kin}} \simeq 10$ K).
\citet{2018arXiv180405205T} found that the G28.28$-$0.36 high-mass star-forming region is a particular cyanopolyyne-rich source with less COMs compared with other sources.
Hence, G28.28$-$0.36 is considered to be a good target region to study the cyanopolyyne chemistry around massive young stellar objects (MYSOs).
Using the Nobeyama 45-m radio telescope, \citet{2016ApJ...830..106T} investigated the main formation mechanism of HC$_{3}$N in G28.28$-$0.36 from its $^{13}$C isotopic fractionation.
The reaction of ``C$_{2}$H$_{2}$ + CN" was proposed as the main formation pathway of HC$_{3}$N, which is consistent with the chemical network simulation conducted by \citet{2009MNRAS.394..221C} and the WCCC model \citep{2008ApJ...681...1385}.

\begin{figure}[ht!]
\plotone{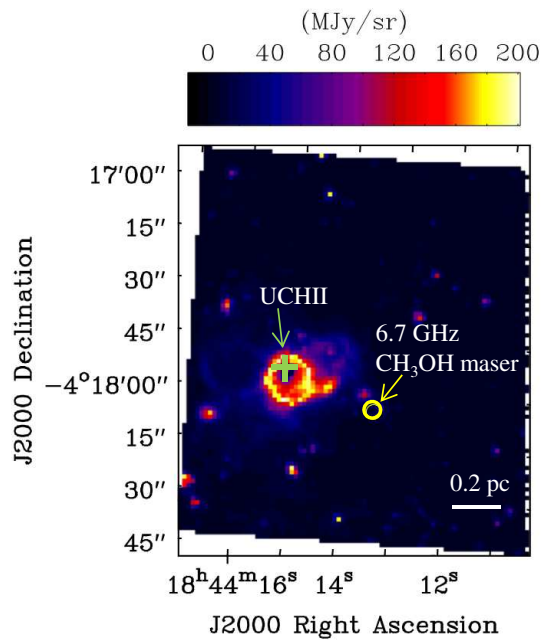}
\caption{Spitzer's IRAC 3.6 $\mu$m image toward G28.28$-$0.36. The open circle and cross indicate the 6.7 GHz methanol maser \citep{2009ApJ...702.1615C} and ultracompact \ion{H}{2} (UC\ion{H}{2}) region \citep{2009A&A...501..539U}, respectively. \label{fig:IR}}
\end{figure}

In this paper, we carried out interferometric observations of cyanopolyynes (HC$_{3}$N, HC$_{5}$N, and HC$_{7}$N) toward the G28.28$-$0.36 high-mass star-forming region ($d = 3$ kpc) with the Karl G. Jansky Very Large Array (VLA).
Figure \ref{fig:IR} shows the Spitzer IRAC 3.6 $\mu$m image\footnote{http://sha.ipac.caltech.edu/applications/Spitzer/SHA/} toward the region.
G28.28$-$0.36 is classified as an Extended Green Object (EGO) source \citep{2008AJ....136.2391C} from the Spitzer Galactic Legacy Infrared Mid-Plane Survey Extraordinaire \citep[GLIMPSE;][]{2003PASP..115..953B}.
In Figure \ref{fig:IR}, the open circle and cross indicate the 6.7 GHz methanol maser \citep{2009ApJ...702.1615C} and ultracompact \ion{H}{2} (UC\ion{H}{2}) region \citep{2009A&A...501..539U}, respectively.
The 6.7 GHz maser is considered to give us the exact position of MYSOs \citep{2013MNRAS.431.1752U}.
A UC\ion{H}{2} region seems to heat the environment.
As shown in Figure \ref{fig:IR}, the ring structure around the UC\ion{H}{2} region is suggestive of expanding motion and on-going massive star formation.
We describe the observational details and data analyses in Section \ref{sec:obs}.
The resultant images and spectra of cyanopolyynes are presented in Section \ref{sec:res}.
We compare the spatial distributions of cyanopolyynes with the infrared images and discuss possible formation mechanisms in Section \ref{sec:dis}.

\section{Observations} \label{sec:obs}

The observations of G28.28$-$0.36 using the VLA Ka-band receiver were carried out in the C configuration with the 27 $\times$ 25-m antennas on March 20th, 2016 (Proposal ID = 16A-084, PI; Kotomi Taniguchi).
The field of view (FoV) is $\sim$ 60\farcs4.
Four spectral windows of the correlator were set at our target lines summarized in Table \ref{tab:line}.
All of these target lines were simultaneously observed.
The channel separation of the correlator is 0.5 km s$^{-1}$.
The angular resolutions and Position Angles (PA) for each line are summarized in Table \ref{tab:line}.

The phase reference center was set at ($\alpha_{2000}$, $\delta_{2000}$) = (18$^{\rm h}$44$^{\rm m}$13\fs3, -04\arcdeg18\arcmin03\farcs0), the 6.7 GHz methanol maser position.
The pointing source is J1832$-$1035 at ($\alpha_{2000}$, $\delta_{2000}$) = (18$^{\rm h}$32$^{\rm m}$20\fs836, $-$10\arcdeg35\arcmin11\farcs2).
The absolute flux density calibration and the bandpass calibration were conducted by observing 3C286 at ($\alpha_{2000}$, $\delta_{2000}$) = (13$^{\rm h}$31$^{\rm m}$08\fs28798, +30\arcdeg30\arcmin32\farcs9589). 
The gain/phase calibration was conducted by observing J1851+0035 at ($\alpha_{2000}$, $\delta_{2000}$) = (18$^{\rm h}$51$^{\rm m}$46\fs7217, +00\arcdeg35\arcmin32\farcs414).

\floattable
\begin{deluxetable}{cccccc}
\tablecaption{Summary of target lines \label{tab:line}}
\tablewidth{0pt}
\tablehead{
\colhead{Species} & \colhead{Transition} & \colhead{Rest Frequency} &\colhead{$E_{\rm {u}}/k$} & \colhead{Angular} & \colhead{PA} \\
\colhead{} & \colhead{} & \colhead{(GHz)} & \colhead{(K)} & \colhead{Resolution} & \colhead{(deg)} 
}
\startdata
HC$_{3}$N & $J=4-3$ & 36.39232 & 4.4 & 0\farcs84 $\times$ 0\farcs63 & -9.92 \\
HC$_{5}$N & $J=14-13$ & 37.276994 & 13.4 & 0\farcs81 $\times$ 0\farcs63 & -11.04 \\
HC$_{7}$N & $J=33-32$ & 37.22349 & 30.4 & 0\farcs82 $\times$ 0\farcs63 & -10.20 \\
CH$_{3}$CN & $J_{\rm K} =2_{0} - 1_{0}$ & 36.7954747 & 2.6 & 0\farcs83 $\times$ 0\farcs64 & -11.22 \\
\enddata
\tablecomments{Rest frequencies are taken from the Cologne Database for Molecular Spectroscopy \citep[CDMS;][]{2005JMoSt...742...215} and the Jet Propulsion Laboratory catalog \citep[JPL catalog;][]{1998JQSRT..60..883P}.}
\end{deluxetable}

We conducted data reduction using the Common Astronomy Software Application \citep[CASA;][]{2007ASPC..376..127M}.
We used the VLA calibration pipeline\footnote{https://science.nrao.edu/facilities/vla/data-processing/pipeline} provided by the National Radio Astronomical Observatory (NRAO) to perform basic flagging and calibration.

The data cubes were imaged using the CLEAN task.
Natural weighting was applied.
The pixel size and image size are $0.2\arcsec$ and $1000 \times 1000$ pixels.
After the CLEAN, we smoothed the cube using the ``imsmooth" command, applying $1\arcsec \times 1\arcsec$ and the position angle of $0\arcdeg$ with the gaussian kernel.
The spatial resolution of $1\arcsec$ of the resultant images corresponds to $\sim 0.015$ pc.
The $1 \sigma$ values are approximately 0.6, 0.7, 0.7, and 0.6 mK for HC$_{3}$N, HC$_{5}$N, HC$_{7}$N, and CH$_{3}$CN, respectively. 
We made the moment zero images of HC$_{3}$N and HC$_{5}$N using the ``immoments" task in CASA.

\section{Results} \label{sec:res}

\begin{figure}[ht!]
\plotone{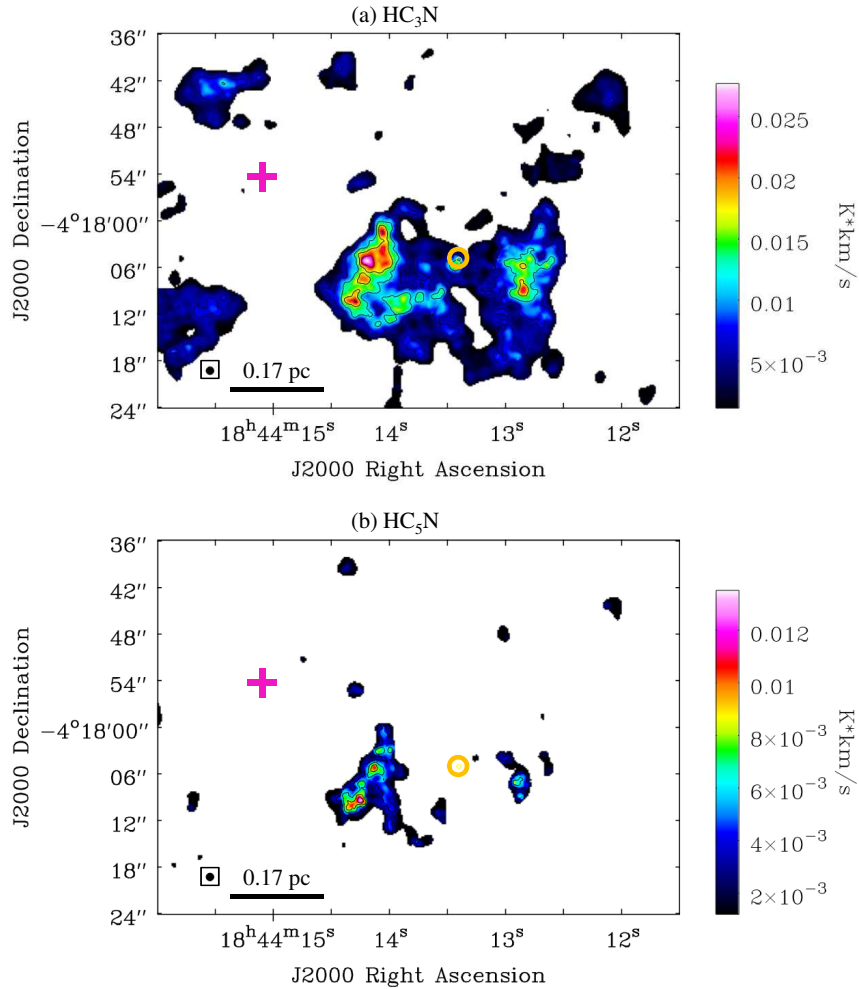}
\caption{Moment zero images of (a) HC$_{3}$N and (b) HC$_{5}$N obtained with the VLA, including the data above $2 \sigma$. The contour levels are 0.2, 0.4, 0.6, and 0.8 of their peak levels, where the peak intensities are 27.9 and 13.5 (mK $\cdot$ km s$^{-1}$) for (a) HC$_{3}$N and (b) HC$_{5}$N, respectively. The rms noise levels are 1.8 and 1.3 (mK $\cdot$ km s$^{-1}$) in the images of HC$_{3}$N and HC$_{5}$N, respectively. The orange open circle and magenta cross indicate the 6.7 GHz methanol maser \citep{2009ApJ...702.1615C} and UC\ion{H}{2} region \citep{2009A&A...501..539U}, respectively. The filled circles at the bottom left corner represent the angular resolution of these images (1\arcsec). \label{fig:mom}}
\end{figure}

Figure \ref{fig:mom} shows the moment zero images of (a) HC$_{3}$N and (b) HC$_{5}$N in G28.28$-$0.36.
The velocity components in the range $V_{\rm {LSR}} = 47.5 - 51.5$ km s$^{-1}$ were integrated in these moment zero images.
The spatial distribution of HC$_{3}$N is more extended than that of HC$_{5}$N, because of their excitation energies of the observed lines (Table \ref{tab:line}).
The observed HC$_{3}$N line has lower excitation energy ($E_{\rm {u}}/k = 4.4$ K) than that of HC$_{5}$N ($E_{\rm {u}}/k = 13.4$ K), and colder envelopes could be traced by HC$_{3}$N.

\begin{figure}[ht!]
\plotone{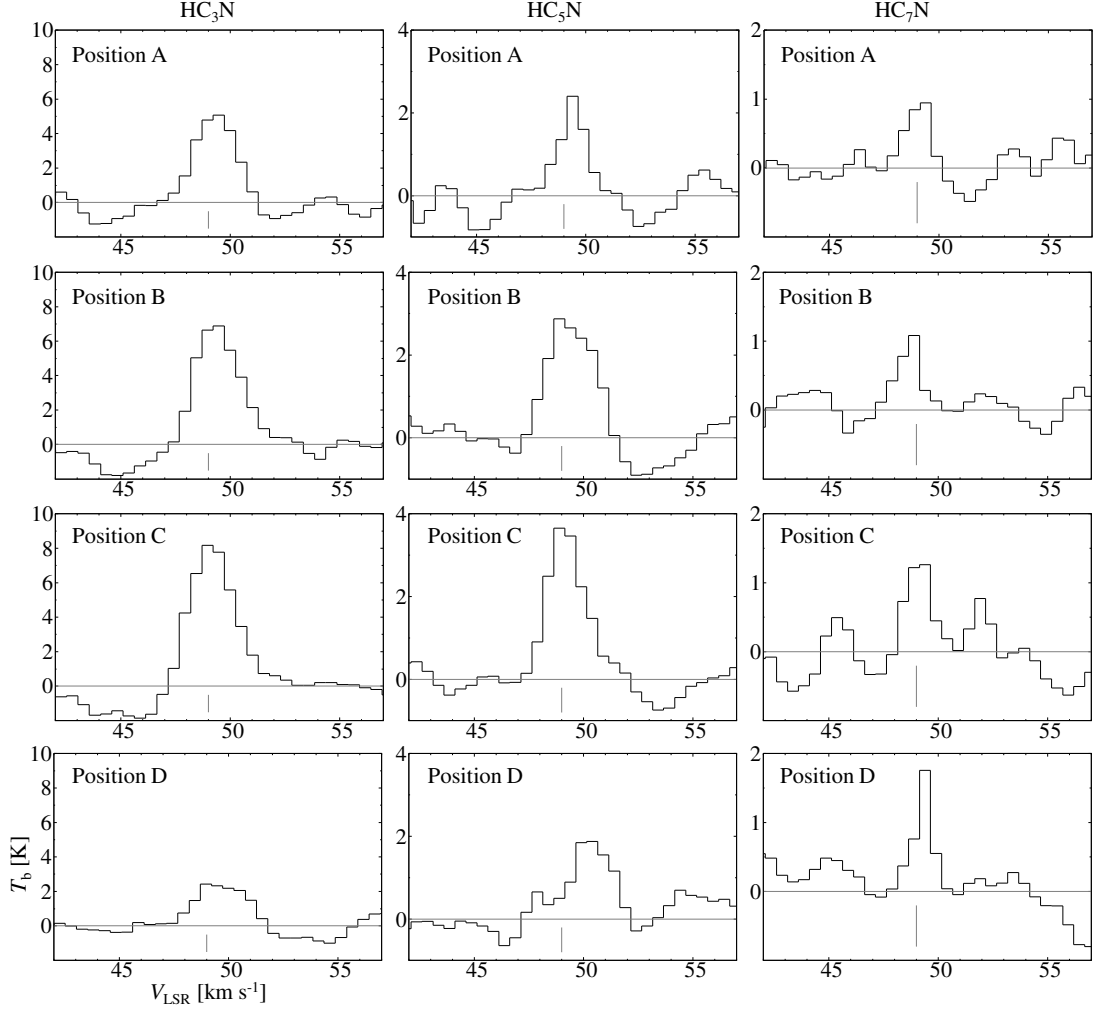}
\caption{HC$_{3}$N, HC$_{5}$N, and HC$_{7}$N spectra at four HC$_{7}$N peak positions labeled as A$-$D, denoting in the panel (b) of Figure \ref{fig:con}. The gray vertical lines show the systemic velocity of the G28.28-0.36 MYSO (49 km s$^{-1}$). The rms noise levels are 0.16, 0.20, and 0.12 K for HC$_{3}$N, HC$_{5}$N, and HC$_{7}$N spectra, respectively. \label{fig:HC7N}}
\end{figure}

Regarding HC$_{7}$N, the signal-to-noise ratio is low, which precludes determination of its spatial distribution.
The bottom panel of Figure \ref{fig:HC7N} shows HC$_{7}$N spectra, as well as HC$_{3}$N and HC$_{5}$N spectra, observed toward its four peak positions A$-$D indicated in the panel (b) of Figure \ref{fig:con}. 
In order to improve the signal-to-noise ratio, we applied the $1 \arcsec$ uvtaper for HC$_{7}$N data.
The intensities of these spectra are estimated within $1.5 \arcsec$ regions, which corresponds to the spatial resolution of HC$_{7}$N with the uvtaper. 
HC$_{7}$N is detected around the regions where HC$_{5}$N is detected as shown in the panel (b) of Figure \ref{fig:mom}.

CH$_{3}$CN emission is undetected at the rms noise level of 0.6 mK.

\section{Discussion} \label{sec:dis}

\subsection{Comparisons of Cyanopolyyne Ratios} \label{sec:d4}

We derived the column densities of HC$_{3}$N, HC$_{5}$N, and HC$_{7}$N at Position A (Figure \ref{fig:HC7N}) assuming the local thermodynamic equilibrium (LTE).
We use the following formulae \citep{1999ApJ...517..209G}: 

\begin{equation} \label{tau}
\tau = - {\mathrm {ln}} \left[1- \frac{T_{\rm b} }{\left\{J(T_{\rm {ex}}) - J(T_{\rm {bg}}) \right\}} \right],  
\end{equation}
where
\begin{equation} \label{tem}
J(T) = \frac{h\nu}{k}\Bigl\{\exp\Bigl(\frac{h\nu}{kT}\Bigr) -1\Bigr\} ^{-1},
\end{equation}  
and
\begin{equation} \label{col}
N = \tau \frac{3h\Delta v}{8\pi ^3}\sqrt{\frac{\pi}{4\mathrm {ln}2}}Q\frac{1}{\mu ^2}\frac{1}{J_{\rm {lower}}+1}\exp\Bigl(\frac{E_{\rm {lower}}}{kT_{\rm {ex}}}\Bigr)\Bigl\{1-\exp\Bigl(-\frac{h\nu }{kT_{\rm {ex}}}\Bigr)\Bigr\} ^{-1}.
\end{equation} 
In Equation (\ref{tau}), $\tau$ and $T_{\rm b}$ denote the optical depth and brightness temperature, respectively.
$T_{\rm{ex}}$ and $T_{\rm {bg}}$ are the excitation temperature and the cosmic microwave background temperature ($\simeq 2.73$ K), respectively.
$J$($T$) in Equation (\ref{tem}) is the effective temperature equivalent to that in the Rayleigh-Jeans law.
In Equation (\ref{col}), {\it N}, $\Delta v$, $Q$, $\mu$, and $E_{\rm {lower}}$ denote the column density, line width (FWHM), partition function, permanent electric dipole moment, and energy of the lower rotational energy level, respectively. 
The brightness temperatures and line widths are obtained by the gaussian fitting of spectra. 
Figure \ref{fig:gaus} in Appendix \ref{sec:a1} shows the fitting results for each spectra, and the obtained spectral line parameters and permanent electric dipole moments of each species are summarized in Table \ref{tab:gaus}.

We derived the column densities assuming the excitation temperatures of 15, 20, 30, and 50 K, respectively.
The column densities of cyanopolyynes and the HC$_{3}$N : HC$_{5}$N : HC$_{7}$N ratios at each excitation temperature are summarized in Table \ref{tab:col}.
The uncertainties in the excitation temperatures do not significantly affect the derived column densities of HC$_{5}$N and HC$_{7}$N, and their derived column densities agree with each other within their the standard deviation errors.
On the other hand, the uncertainties in the excitation temperatures bring larger differences in the HC$_{3}$N column density.
\floattable
\begin{deluxetable}{ccccc}
\tablecaption{Column densities of cyanopolyynes at Position A \label{tab:col}}
\tablewidth{0pt}
\tablehead{
\colhead{$T_{\rm {ex}}$} & \colhead{$N$(HC$_{3}$N)} & \colhead{$N$(HC$_{5}$N)} & \colhead{$N$(HC$_{7}$N)} & \colhead{HC$_{3}$N : HC$_{5}$N : HC$_{7}$N} \\
\colhead{(K)} & \colhead{($\times 10^{14}$ cm$^{-2}$)} & \colhead{($\times 10^{13}$ cm$^{-2}$)} & \colhead{($\times 10^{13}$ cm$^{-2}$)} & \colhead{}
}
\startdata
15 & $1.36 \pm 0.14$ & $4.4 \pm 0.9$ & $4.0 \pm 1.2$ & 1.00 : 0.32 : 0.29 \\
20 & $1.47 \pm 0.15$ & $4.3 \pm 0.9$ & $3.1 \pm 0.9$ & 1.00 : 0.29 : 0.21 \\ 
30 & $1.84 \pm 0.19$ & $4.9 \pm 1.1$ & $2.6 \pm 0.8$ & 1.00 : 0.26 : 0.14 \\
50 & $2.7 \pm 0.3$ & $6.5 \pm 1.4$ & $2.8 \pm 0.8$ & 1.00 : 0.24 : 0.11 \\
\enddata
\tablecomments{The errors represent the standard deviation. The errors of column densities are derived from uncertainties of the gaussian fitting (see Table \ref{tab:gaus} in Appendix).}
\end{deluxetable}

The HC$_{3}$N : HC$_{5}$N : HC$_{7}$N ratios at Position A are derived to be 1.0 : $\sim 0.3$ : $\sim 0.2$.
The ratios in L1527, which is one of the WCCC sources, are 1.0 : $0.3-0.6$ : $\sim 0.1$ \citep{2008ApJ...672..371S, 2009ApJ...702.1025S}.
The tendency of the ratios at Position A seems to be similar to that in L1527.

\subsection{Comparison of Spatial Distributions between Cyanopolyynes and 450 $\mu$m Dust Continuum} \label{sec:d1}

\begin{figure}[ht!]
\plotone{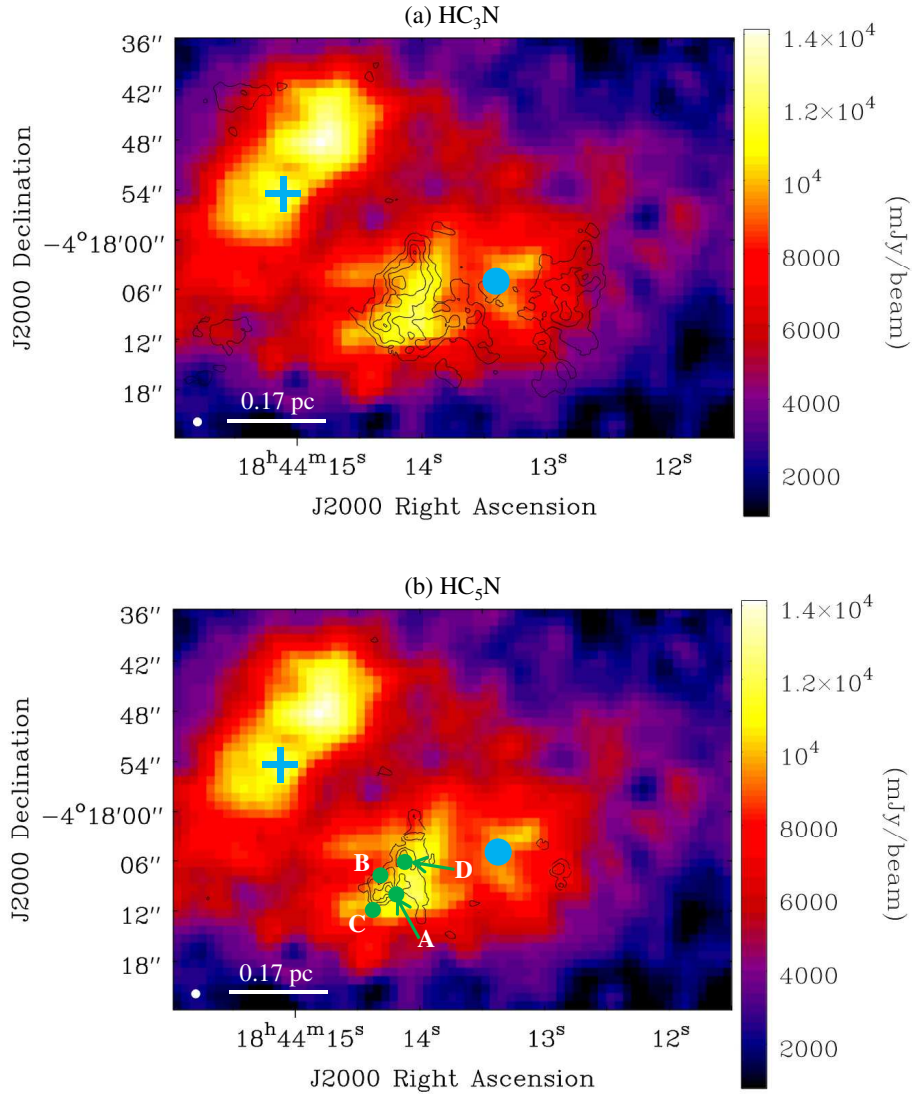}
\caption{450 $\mu$m continuum image obtained with the SCUBA installed on the JCMT overlaid by black contours of (a) HC$_{3}$N and (b) HC$_{5}$N moment zero images, the same ones as in Figure \ref{fig:mom}. The contour levels are 0.2, 0.4, 0.6, and 0.8 of their peak intensities 27.9 and 13.5 (mK $\cdot$ km s$^{-1}$) for (a) HC$_{3}$N and (b) HC$_{5}$N, respectively. The blue filled circle and blue cross indicate the 6.7 GHz methanol maser and UC\ion{H}{2} region, respectively. The A--D positions are indicated as the green filled circles. The filed white circles at the bottom left corner represent the angular resolution of the images of HC$_{3}$N and HC$_{5}$N (1\arcsec). \label{fig:con}}
\end{figure}

Figure \ref{fig:con} shows 450 $\mu$m dust continuum images overlaid by the black contours of moment zero images of (a) HC$_{3}$N and (b) HC$_{5}$N, respectively.
The 450 $\mu$m data, which are available from the James Clerk Maxwell Telescope (JCMT) Science Archive\footnote{http://www.cadc-ccda.hia-iha.nrc-cnrc.gc.ca/en/jcmt/index.html}, were obtained with the SCUBA installed on the JCMT.
The main beam size of the SCUBA is $7.9\arcsec$ at 450 $\mu$m, corresponding to $\sim 0.11$ pc.

Three strong 450 $\mu$m continuum emission peaks can be recognized; the UC\ion{H}{2} position, the north-west position from the UC\ion{H}{2}, and the east position from the 6.7 GHz methanol maser.
The spatial distribution of HC$_{5}$N is consistent with the 450 $\mu$m continuum peak of the east position from the 6.7 GHz methanol maser (hereafter Cyanopolyyne-rich clump, panel (b) of Figure \ref{fig:con}).
The spatial distribution of HC$_{3}$N seems to surround the Cyanopolyyne-rich clump (panel (a) of Figure \ref{fig:con}), not only at the Cyanopolyyne-rich clump.
HC$_{3}$N emission is also located at the west position of the 6.7 GHz methanol maser position, or the edge of the 450 $\mu$m continuum.
Small HC$_{5}$N emission region is seen at the same edge of the 450 $\mu$m continuum.
We will briefly discuss a possible origin of this emission region in Section \ref{sec:d2}.

\subsection{Possible Nature of the 450 $\mu$m Continuum Peak Position associated with Cyanopolyynes} \label{sec:d2}

\begin{figure}[ht!]
\plotone{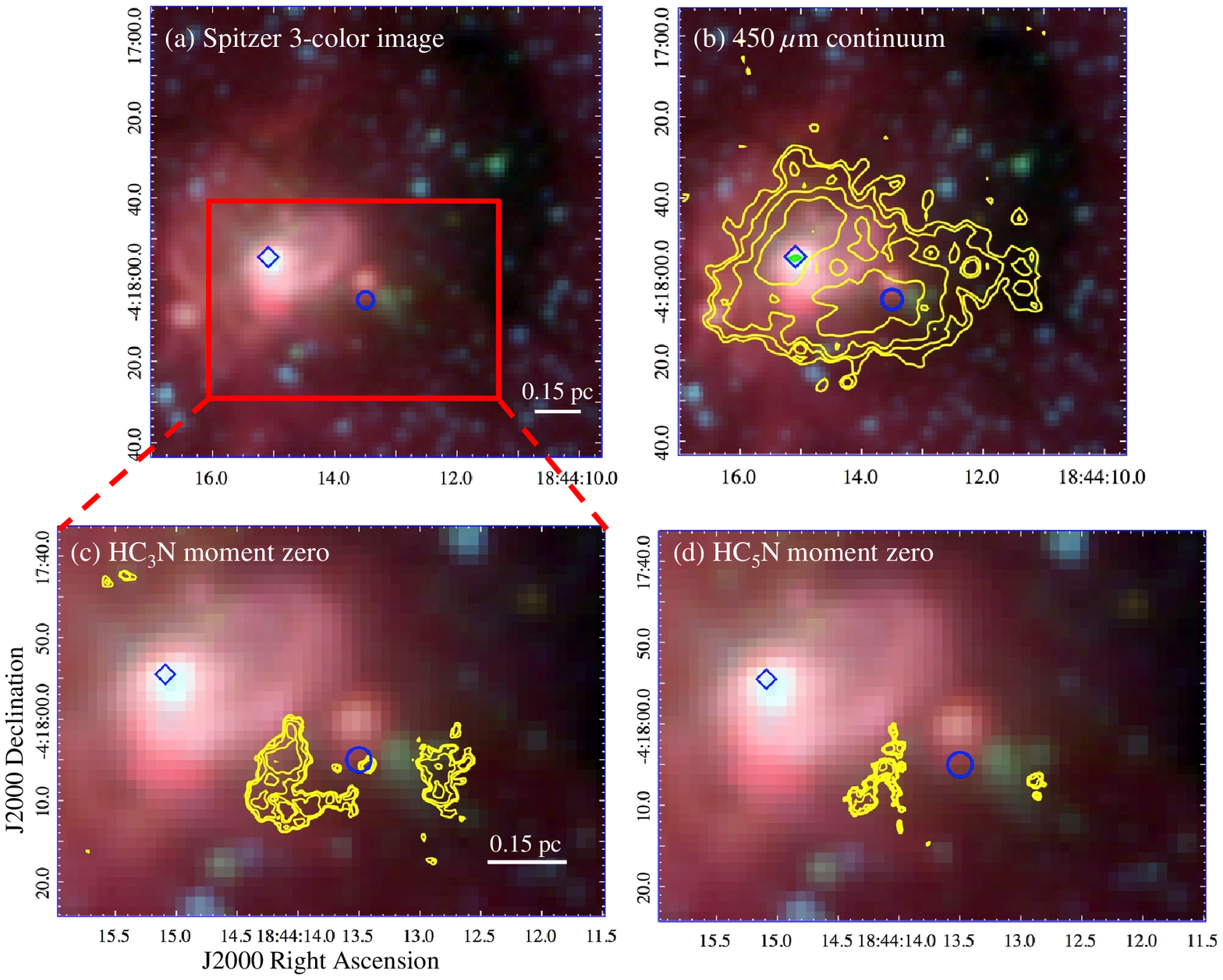}
\caption{(a) Spitzer 3-color image (red; 8.0 $\mu$m, green; 4.5 $\mu$m, blue; 3.6 $\mu$m), overlaid by yellow contours showing (b) 450 $\mu$m continuum, (c) HC$_{3}$N moment zero image, and (d) HC$_{5}$N moment zero image. The green contour in panel (b) is the 8.3 mm continuum emission obtained with the VLA. The contour levels are 90, 80, 70, 60, 50 \% of their peak levels. The blue open circle and diamond indicate the 6.7 GHz methanol maser and UC\ion{H}{2} region, respectively. \label{fig:rgb}}
\end{figure}

We examine four possible types of objects of the Cyanopolyyne-rich clump (Section \ref{sec:d1}); a hot core, a starless clump, low- or intermediate-mass protostellar core(s), and a photo-dominated region (PDR) driven by the associated UC\ion{H}{2} region.

Figure \ref{fig:rgb} shows the Spitzer 3-color (3.6 $\mu$m, 4.5 $\mu$m, 8.0 $\mu$m) images\footnote{http://atlasgal.mpifr-bonn.mpg.de/cgi-bin/ATLASGAL$_{-}$DATABASE.cgi}, overlaid by yellow contours showing (b) 450 $\mu$m continuum, (c) HC$_{3}$N moment zero image, and (d) HC$_{5}$N moment zero image, respectively.
In panel (b), the green contours show the 8.3 mm continuum emission obtained simultaneously with cyanopolyynes by the VLA.
The 8.3 mm continuum peak is compact and well consistent with the UC\ion{H}{2} region.
At the Cyanopolyyne-rich clump, no point source can be recognized from the Spitzer image, which suggests that no massive young protostar is currently present at the clump position.
Therefore, the possibility of hot core is unrealistic. 

We derived the average column density of H$_{2}$, $N$(H$_{2}$), of the Cyanopolyyne-rich clump from the 450 $\mu$m continuum data using the following formula \citep{2005ApJ...632..982S}:
\begin{equation} \label{H2}
N({\rm {H}}_2) = 2.02 \times 10^{20} {\rm {cm}}^{-2} \Bigl(e^{1.439(\lambda /{\rm {mm}})^{-1}(T/10 {\rm {K}})^{-1}}-1\Bigr)\Bigl(\frac{\kappa_{\nu}}{0.01\: {\rm {cm}}^{2}\: {\rm {g}}^{-1}}\Bigr)^{-1}\Bigl(\frac{S_{\nu}^{\rm {beam}}}{{\rm {mJy}}\: {\rm {beam}}^{-1}}\Bigr)\Bigl(\frac{\theta_{\rm {HPBW}}}{10\: {\rm {arcsec}}}\Bigr)^{-2}\Bigl(\frac{\lambda}{{\rm {mm}}}\Bigr)^{3}.
\end{equation}
We estimated the continuum flux ($S_{\nu}^{\rm {beam}}$) toward the Cyanopolyyne-rich clump within a size of $14\arcsec$, and the flux intensity is $9.7 \times 10^{3}$ mJy beam$^{-1}$.
We assumed that $\kappa_{\nu}$ is 0.0619 cm$^{2}$ g$^{-1}$ at $\lambda = 450$ $\mu$m.
$T$ is the dust temperature, and we assumed it to be $30 \pm 15$ K.
The derived $N$(H$_{2}$) value is ($2.8^{+8.1}_{-1.3}$)$\times 10^{22}$ cm$^{-2}$, which is similar to the typical value of massive clumps forming young star clusters \citep[e.g.,][]{2018ApJ...855...45S}.
The value is an average for the clump and then it can be considered as the lower limits for cores.
The derived $N$(H$_{2}$) value for the Cyanopolyyne-rich clump is higher than the threshold value for star formation cores, $N$(H$_{2}$) $\approx 9 \times 10^{21}$ cm$^{-2}$ \citep[e.g.,][]{2002A&A...385..909T}.
Hence, the Cyanopolyyne-rich clump is considered to contain deeply embedded low- or intermediate-mass protostellar core(s), and is not a starless clump.

In case that a low- or intermediate-mass protostar has been born, the WCCC mechanism is likely to work; cyanopolyynes are formed from CH$_{4}$ evaporated from grain mantles in the lukewarm gas ($T = 20-30$ K). 
This may be supported by similar HC$_{3}$N : HC$_{5}$N : HC$_{7}$N ratios between this clump and L1527 (Section \ref{sec:d4}).

In order to investigate the possibility of the PDR, we compare the fractional abundances of cyanopolyynes at the Cyanopolyyne-rich clump with those derived from a PDR chemical network simulation \citep{2017AA...605A..88L}.
We derived the fractional abundances, $X$($a$)$= N$($a$)$/N$(H$_{2}$), of cyanopolyynes at Position A as summarized in Table \ref{tab:frac}.
Recent chemical network simulation about the Horsehead nebula, which is one of the most studied PDRs, estimated the HC$_{3}$N and HC$_{5}$N abundances \citep{2017AA...605A..88L}.
The PDR and Core positions are positions with a visual extinctions ($A_{v}$) of $\sim 2$ mag and $\sim 10-20$ mag, respectively.
We summarize the fractional abundances of these two positions in Table \ref{tab:frac}.
The $X$(HC$_{3}$N) and $X$(HC$_{5}$N) values at the Cyanopolyyne-rich clump are significantly higher than those in PDR position by three order of magnitudes and more than four order of magnitudes, respectively.
Moreover, the HC$_{3}$N and HC$_{5}$N fractional abundances at the Cyanopolyyne-rich clump are still higher than those in Core position in the PDR model by a factor of $\sim 55$ and $\sim 80$, respectively.
Therefore, cyanopolyynes at the Cyanopolyyne-rich clump in the G28.28$-$0.36 high-mass star-forming region are not formed by the PDR chemistry.
If we take large uncertainties in deriving $N$(H$_{2}$) values into consideration, the differences between the Cyanopolyyne-rich clump and the PDR models are plausible.
However, we cannot completely exclude any effects from the UCHII region such as UV radiation.

\floattable
\begin{deluxetable}{lccccc}
\tablecaption{Comparison of fractional abundances of cyanopolyynes \label{tab:frac}}
\tablewidth{0pt}
\tablehead{
\colhead{Source} & \colhead{Method} & \colhead{$X$(HC$_{3}$N)} & \colhead{$X$(HC$_{5}$N)} & \colhead{$X$(HC$_{7}$N)} &\colhead{Reference\tablenotemark{a}} \\
\colhead{ } &\colhead{} & \colhead{($\times 10^{-9}$)} & \colhead{($\times 10^{-9}$)} & \colhead{($\times 10^{-9}$)} & \colhead{ }
}
\startdata
Cyanopolyyne-rich clump & Observation & $6.6^{+5.5}_{-4.9}$ & $1.7^{+1.5}_{-1.3}$ & $0.9^{+0.8}_{-0.7}$ & 1 \\
Horsehead nebula (PDR position) & Simulation  & $3.1 \times 10^{-3}$ & $4.3 \times 10^{-5}$ & ... & 2 \\
Horsehead nebula (Core position) & Simulation  & 0.12 & 0.022 & ... & 2 \\
L1527 (WCCC) & Observation & 0.43 -- 0.96 & 0.05 -- 0.24 & $\sim 0.05$ & 3,4,5 \\
\enddata
\tablecomments{The errors represent the standard deviation.}
\tablenotetext{a}{(1) This work; (2) \citet{2017AA...605A..88L}; (3) \citet{2002AA...389..908J}; (4) \citet{2008ApJ...672..371S}; (5) \citet{2009ApJ...702.1025S} }
\end{deluxetable}

We also compare the fractional abundances of HC$_{3}$N, HC$_{5}$N, and HC$_{7}$N at the Cyanopolyyne-rich clump to those in L1527.
These fractional abundances at the Cyanopolyyne-rich clump are higher than those in L1527 by a factor of $7-15$, $7-37$, and $\sim 17$, respectively.
Hence, the cyanopolyynes at the Cyanopolyyne-rich clump seem to be more abundant compared to L1527, even if we take uncertainties in fractional abundances into account.
It cannot be excluded that several low-mass protostars are concentrated within a small region (e.g., $\sim 0.02$ pc corresponding to $1.5\arcsec$ at the distance of 3 kpc). 
As another interpretation, these results may imply that there is an intermediate-mass protostar because an intermediate-mass protostar heats its wider surroundings compared to a low-mass protostar; the size of lukewarm envelopes will be larger and the column densities of carbon-chain species in such lukewarm envelopes will increase.
This could explain the results that the cyanopolyynes at the Cyanopolyyne-rich clump are more abundant than those in L1527.

\begin{figure}[ht!]
\plotone{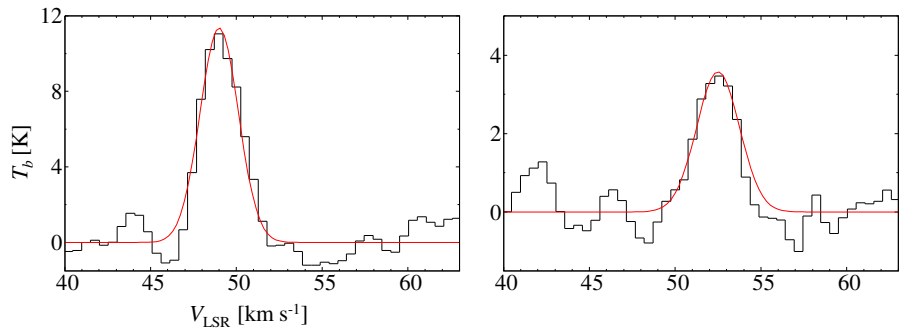}
\caption{Spectra of HC$_{3}$N and HC$_{5}$N at the shock region. The red lines show the Gaussian fitting results. \label{fig:outflow}}
\end{figure}

In Figure \ref{fig:rgb} (d), compact HC$_{5}$N emission region is located at the west position of the 6.7 GHz methanol maser position.
This position corresponds to the edge of the 4.5 $\mu$m emission, which seems to trace shock regions \citep{2008AJ....136.2391C}.
Figure \ref{fig:outflow} shows spectra of HC$_{3}$N and HC$_{5}$N at the shock region.
The line widths (FWHM) derived from the Gaussian fit are $2.7 \pm 0.2$ and $2.6 \pm 0.2$ km s$^{-1}$ for HC$_{3}$N and HC$_{5}$N, respectively.
These line widths at the shock region are wider those at the Cyanopolyyne-rich clump, $2.05 \pm 0.16$ and $1.4 \pm 0.2$ km s$^{-1}$ for HC$_{3}$N and HC$_{5}$N, respectively.   
These results probably support that the cyanopolyynes are the shock origin.
Molecules evaporated from grain mantles such as C$_{2}$H$_{2}$ may be parent species of the cyanopolyynes.
It is still unclear whether HC$_{5}$N is formed and can survive in shock regions, but this may be the first observational result showing that HC$_{5}$N can be enhanced in shock regions.

\subsection{Comparison with Previous Single-Dish Telescope Observations} \label{sec:d3}

\citet{2017ApJ...844...68T} reported the detection of high-excitation-energy ($E_{\rm {u}}/k \simeq 100$ K) lines of HC$_{5}$N with the Nobeyama 45-m radio telescope (HPBW = 18\arcsec) toward the methanol maser position, which is considered to be a MYSO position \citep{2013MNRAS.431.1752U}.
The significant HC$_{5}$N peak is not seen at the methanol maser position in the VLA map.
This is probably caused by the different excitation energies of the observed lines.
The lines observed with the VLA have low excitation energies (Table \ref{tab:line}), and preferably trace lower-temperature regions such as low- or intermediate-mass protostellar cores or envelopes.
On the other hand, HC$_{5}$N emission observed with the Nobeyama 45-m telescope seems to come from higher-temperature regions closer to the methanol maser or the MYSO.
In these higher-temperature regions, HC$_{5}$N should be highly excited and low excitation energy lines observed with the VLA should be weak.
In contrast, the hotter components detected with the Nobeyama 45-m telescope were not detected with the VLA, because the low-excitation-energy lines are not suitable tracers of the hot components.
Combining with the VLA and Nobeyama 45-m telescope results, both the MYSO associated with the methanol maser and low- or intermediate-mass protostellar core(s) at the Cyanopolyyne-rich clump appear to be rich in cyanopolyynes.

It is still unclear why the G28.28$-$0.36 high-mass star-forming region is a cyanopolyyne-rich/COMs-poor source \citep{2018arXiv180405205T}.
Studies about chemical diversity among high-mass star-forming regions will become a key to our understanding of massive star formation processes.
We need the high-spatial-resolution and higher-frequency band observations in order to investigate the spatial resolution of higher temperature components of cyanopolyynes to confirm that cyanopolyynes exist at the hot core position. 

\section{Conclusions} \label{sec:con}

We have carried out interferometric observations of cyanopolyynes (HC$_{3}$N, HC$_{5}$N, and HC$_{7}$N) toward the G28.28$-$0.36 high-mass star-forming region with the VLA Ka-band.
We obtained the moment zero images of HC$_{3}$N and HC$_{5}$N and tentatively detected HC$_{7}$N.
The spatial distributions of HC$_{3}$N and HC$_{5}$N are consistent with the 450 $\mu$m dust continuum clump, i.e., the Cyanopolyyne-rich clump.
The HC$_{3}$N : HC$_{5}$N : HC$_{7}$N column density ratios are estimated at 1.0 : $\sim 0.3$ : $\sim 0.2$ at Position A.
The Cyanopolyyne-rich clump seems to contain deeply embedded low- or intermediate-mass protostellar core(s).
The most probable formation mechanism of the cyanopolyynes at the Cyanopolyyne-rich clump is the WCCC mechanism.
We possibly found the HC$_{3}$N and HC$_{5}$N emission in the shock region.





\acknowledgments

We express our sincere thanks and appreciate for the staff of the National Radio Astronomy Observatory.
The National Radio Astronomy Observatory is a facility of the National Science Foundation operated under cooperative agreement by Associated Universities, Inc.
KT appreciates support from a Granting-Aid for Science Research of Japan (17J03516).

%

\vspace{5mm}
\facilities{Karl G. Jansky Very Large Array (VLA)}


\software{Common Astronomy Software Applications (CASA)}



\appendix
\section{Gaussian Fitting Results and Spectral Line Parameters} \label{sec:a1}

\begin{figure}[ht!]
\plotone{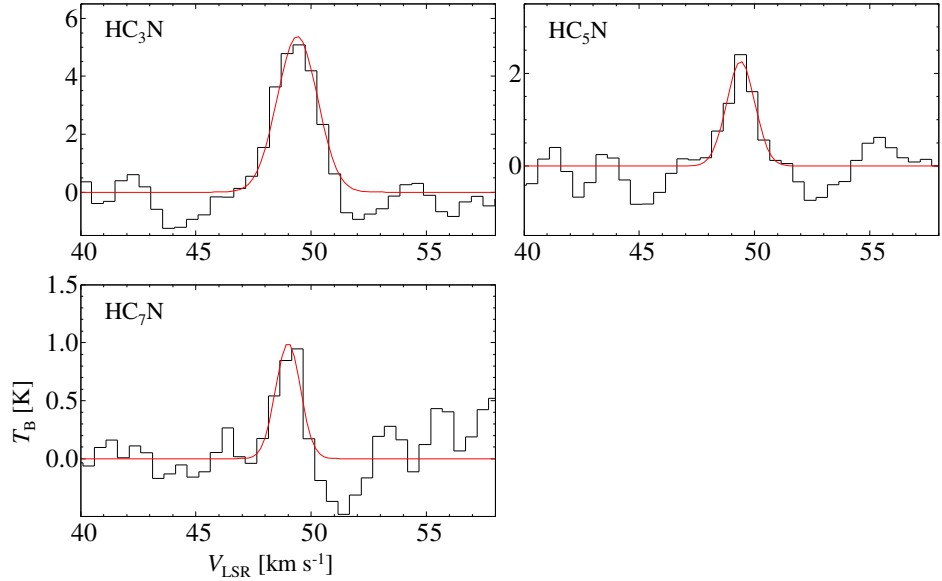}
\caption{Spectra of HC$_{3}$N, HC$_{5}$N, and HC$_{7}$N at Position A obtained with the VLA. The best gaussian fits are shown overlaid in red. \label{fig:gaus}}
\end{figure}

We show the gaussian fitting results for the spectra of HC$_{3}$N, HC$_{5}$N, and HC$_{7}$N at Position A in Figure \ref{fig:gaus}.
The obtained spectral line parameters are summarized in Table \ref{tab:gaus}.

\floattable
\begin{deluxetable}{ccccc}
\tablecaption{Spectral line parameters at Position A \label{tab:gaus}}
\tablewidth{0pt}
\tablehead{
\colhead{Species} & \colhead{$\mu$} & \colhead{$T_{\rm {b}}$} & \colhead{FWHM} & \colhead{$\int T_{\mathrm {b}}dv$} \\
\colhead{} & \colhead{(Debye)} & \colhead{(K)} & \colhead{(km s$^{-1}$)} & \colhead{(K km s$^{-1}$)}
}
\startdata
HC$_{3}$N & 3.73 & $5.4 \pm 0.4$ & $2.05 \pm 0.16$ & $11.7 \pm 1.2$ \\
HC$_{5}$N & 4.33 & $2.3 \pm 0.3$ & $1.4 \pm 0.2$ & $3.5 \pm 0.8$ \\
HC$_{7}$N & 4.82 & $1.0 \pm 0.2$ & $1.3 \pm 0.3$ & $1.3 \pm 0.4$ \\
\enddata
\tablecomments{The errors represent the standard deviation. Values of their permanent electric dipole moments are taken from the Cologne Database for Molecular Spectroscopy \citep[CDMS;][]{2005JMoSt...742...215}.}
\end{deluxetable}

\end{document}